\documentstyle{article}

\begin{document}
\baselineskip=18pt
\title{\hspace{11cm}{\small\bf IMPNWU-960810}\\
\vspace{2cm}
The nondynamical r-matrix structure of  the elliptic Calogero-Moser
Model}

\author{
 Bo-yu Hou$^{b}$ and Wen-li Yang$^{a,b}$ 
\thanks{e-mail :wlyang@nwu.edu.cn}
\thanks{Fax    :0086-029-8303511}
\\
\bigskip\\
$^{a}$ CCAST ( World Laboratory ), P.O.Box 8730 ,Beijing 100080, China\\
$^{b}$ Institute of Modern Physics, Northwest 
University, Xian 710069, China
\thanks{Mailing address}}

\maketitle

\begin{abstract}
\footnote{This paper was given as a lecture in the international meeting ``
New developments in quantum field theories"\\ 
in Urumqi,China,10-19 August,1996}

In this paper, we construct a new Lax operator for the elliptic
Calogero-Moser model with N=2. The nondynamical r-matrix structure 
of this Lax operator is also studied . 
The relation between our Lax operator and the Lax operator 
given by Krichever is also obtained.

{\bf Mathematics Subject Classification : }70F10 , 70H33 , 81U10.
\end{abstract}
\section{Introduction}
The elliptic Calogero-Moser (CM) model[1-4] is the system  
of N one-dimensional particles interacting by two-particle potential 
of the elliptic type. It is well-known that the CM model is completely 
integrable[1-8]. The Lax operator of this system , which is the most   
effective way to construct the complete set of integrals of motion, was 
found by Krichever[6] . The classical r-matrix structure 
of  the Lax operator given by Krichever for the CM model was obtained 
by Sklyanin[7]. This r-matrix is a natural generalization of the 
 matrix found by Avan at al [9] for the trigonometric potential. 
 There exists a specific feature that the r-matrix for these model turns 
 out to be of dynamical  type (i.e it depends on the dynamical variables) 
 and satisfy a generalized Yang-Baxter equation[7-9]. Very recently 
 development was connected with the geometrical interpretation of the 
 dynamical r-matrix in terms of the Hamiltonian reduction[10].

The difficulty presented by the dynamic aspect of the r-matrix is : {\bf I.} 
the Poisson algebra of a model ,whose structural constants are given by a 
dynamical r-matrix, is generally speaking no longer closed; {\bf II.} 
To solve the quantization problem is still an open problem. So far ,only  
for one particular example---the spin generalization of the CM model---a 
proper algebraic setting (the Gervais-Neveu-Felder equation ) was found
[11] which also allows to quantize the model. To overcome the above 
difficulties may be that whether another Lax operator for the CM model which 
has a numberical r-matrix structure could be found. In this paper, we 
construct a new Lax operator for the CM model with N=2. In this new Lax 
representation , we find its r-matrix being of numerical type , which 
satisfies the classical Yang-Baxter equation. Further, the relation between 
the old one (given by Krichever) and ours is also obtained.

\section{Review of the CM model}
The Calogero-Moser model is the system of N one-dimensional particles 
interactin by the two-particle potential
\begin{eqnarray}
& & V(q_{ij})=Q(q_{ij})\ \ \ ,\ \ q_{ij}=q_{i}-q_{j}\ \ ,\ \ i,j=1,....N\ \ ,\\
& &Q(q)-Q(u)=E(u,q)E(u,-q)\nonumber
\end{eqnarray}
\noindent and $E(u,q)$ is an elliptic function which is defined in Eq.(3). 
The identity Eq.(3a) is used in the second equation. 
In this paper, we restrict ourselves to the case N=2.

In terms of the canonical variables $p_{i}\ \ , q_{i}$ (i=1,2)
\begin{eqnarray*}
\{p_{i},p_{j}\}=0\ \ ,\ \ \{q_{i},q_{j}\}=0\ \ ,\ \ \{p_{i},q_{j}\}
=\delta_{i,j}\ \ ,
\end{eqnarray*}
\noindent the Hamiltonian of the system is expressed as 
\begin{eqnarray}
H=\sum_{i=1}^{2}p_{i}^{2} +\sum_{i\ne j}V(q_{ij}).
\end{eqnarray}
\noindent The Hamiltonian Eq.(2) with the potential Eq.(1) is known to be  
completely integrable. The most effective way to show its integrability  is 
to construct the Lax representation for the system (namely, to find the Lax 
operator). One  Lax representation for the CM model was first found by   
Krichever[6]
\begin{eqnarray}
\stackrel{\sim }{L}(u)=\left(\begin{array}{ll}
p_{1}&E(u,q_{12})
\\E(u,q_{21})&p_{2}
\end{array}\right)
\ \ ,\ \ E(u,q)=\frac{\sigma(u+q)}{\sigma(u)\sigma(q)}\ \ ,
\end{eqnarray}
\noindent where  the elliptic function $\sigma(u)$ is defined in Eq.(6).
The Hamiltonian in Eq.(2) can be written as 
\begin{eqnarray*}
H=tr(\stackrel{\sim}{L}^{2}(u))+V(u)
\end{eqnarray*}
\noindent where $V(u)$ does not depend upon the dynamical variables and  
the identity
$$
E(u,q_{12})E(u,q_{21})=V(q_{12})-V(u)\eqno(3a)
$$
\noindent is used. The motion equation can be written in the following form
[6]
\begin{eqnarray*}
\frac{d}{dt}\stackrel{\sim }{L}(u)=\{\stackrel{\sim }{L}(u),H\}
=[\stackrel{\sim }{L}(u),\stackrel{\sim }{M}(u)]
\end{eqnarray*}
\noindent The Poisson bracket of the Lax operator $\stackrel{\sim}{L}(u)$ 
can be described in terms of r-matrix form[7]
\begin{eqnarray}
& &\{\stackrel{\sim }{L}_{1}(u),\stackrel{\sim }{L}_{2}(v)\}
=[\stackrel{\sim }{r}_{12}(u,v),\stackrel{\sim }{L}_{1}(u)]
-[\stackrel{\sim }{r}_{21}(v,u),\stackrel{\sim }{L}_{2}(v)]
\end{eqnarray}
\begin{eqnarray*}
& &\stackrel{\sim }{r}_{12}(u,v)=\left(\begin{array}{llll}a&d_{12}&&\\
d_{21}& &c_{12}&\\&c_{21}&&d_{12}\\ &&d_{21}&a\end{array}\right)\\
& &a=-\xi (u-v)-\xi (v)\ \ ,\ \ \xi (u)=\partial _{u}\{ln\sigma(u)\}\\
& &c_{ij}=-E(u-v,q_{ij})\ \,\ \ d_{ij}=-\frac{1}{2}E(v,q_{ij})
\end{eqnarray*}
It can be seen that the classical r-matrix $\stackrel{\sim}{r}(u,v)$ is a 
dynamical r-matrix (i.e the matrix element of $\stackrel{\sim}{r}(u,v)$ 
depends upon the dynamical variables $q_{i}$ ). The Poisson bracket 
Eq.(4) leads to the evolution integrals $tr(\stackrel{\sim}{L}^{n}(u))$ 
of the motion. Sklyanin also shown that the dynamical r-matrix 
$\stackrel{\sim}{r}(u,v)$ defined by Eq.(4) satisfies the generalized 
Yang-Baxter equation[7]
\begin{eqnarray}
& &[R^{(123)},\stackrel{\sim}{L}^{(1)}]+[R^{(231)},\stackrel{\sim}{L}^{(2)}]
+[R^{(312)},\stackrel{\sim}{L}^{(3)}]=0
\end{eqnarray}
\noindent where 
\begin{eqnarray*}
R^{(123)}\equiv \stackrel{\sim}{r}_{(123)}-
\{\stackrel{\sim}{r}_{13},\stackrel{\sim}{L}^{(2)}\}
+\{\stackrel{\sim}{r}_{12},\stackrel{\sim}{L}^{(3)}\},
\end{eqnarray*}
\noindent and 
\begin{eqnarray*}
\stackrel{\sim}{r}_{(123)}=[\stackrel{\sim}{r}_{12},\stackrel{\sim}{r}_{13}]
+[\stackrel{\sim}{r}_{12},\stackrel{\sim}{r}_{23}]-
[\stackrel{\sim}{r}_{13},\stackrel{\sim}{r}_{32}]
\end{eqnarray*}
Due to the r-matrix $\stackrel{\sim}{r}(u,v)$ depending on the dynamical 
variables, the Poisson bracket of $\stackrel{\sim}{L}(u)$ is no longer 
closed.The quantum version of Eq.(4) and the generalized Yang-Baxter 
equation is still not found except the spin generalization of the CM model  
in which the Gervais-Neveu-Felder equation was found[11].

\section{ The new Lax representation for CM model}
The Lax represenation $\stackrel{\sim}{L}(u)$ of the CM model given by 
Krichever in Eq.(3) and its classical r-matrix $\stackrel{\sim}{r}(u,v)$ 
given by Sklyanin in Eq.(4) leads to some difficulty[7] in the 
investigation of the CM model : the Poisson algebra of the Lax operator  
is no longer closed and the quantum version of Eq.(4) is still not found.
 This motivate us to find a new Lax representation of the CM model. 
 Fortunately, we find the new Lax representation of the CM model, in which 
  the classical r-matrix is numerical one (This kind Lax representation 
  does not always exist for generic system with dynamical r-matrix). 
  In this section, we construct this new lax operator (in contrast to 
the Lax operator given by Krichever, we call the Lax operator found by us  
as the new Lax operator).

First,let us define some elliptic functions
\begin{eqnarray}
& &\theta^{(j)}(u)=
\theta\left[\begin{array}{c}\frac{1}{2}-\frac{j}{2}\\ 
\frac{1}{2}\end{array}\right](u,2\tau)\nonumber\\
& &\sigma(u)=\theta\left[\begin{array}{c}\frac{1}{2}\\ 
\frac{1}{2}\end{array}\right](u,\tau)\\
& &\theta\left[\begin{array}{c}a\\ b\end{array}\right](u,\tau)
=\sum_{m=-\infty}^{\infty}exp\{i\pi[(m+a)^{2}\tau + 2(m+a)(z+b)]\}\nonumber\\
& &\theta'^{(j)}(u)=\partial_{u}\{\theta^{(j)}(u)\}\nonumber
\end{eqnarray}
\noindent where $\tau$ is a complex number with $Im(\tau)>0$ .We find that 
there exist another Lax representation for the CM model and denote it by  
$L(u)$
\begin{eqnarray}
L(u)=\left(\begin{array}{ll}L_{11}(u)&L_{12}(u)\\
L_{21}(u)&L_{22}(u)\end{array}\right),
\end{eqnarray}
\noindent where
\begin{eqnarray*}
L_{11}(u)&=&\{\theta_{1}(u-2\overline{q}_{1})\theta_{2}(u-2\overline{q}_{2})p_{1}
-\theta_{1}(u-2\overline{q}_{2})\theta_{2}(u-2\overline{q}_{2})E(u,q_{21})\\
& &\ \ +\theta_{1}(u-2\overline{q}_{1})\theta_{2}(u-2\overline{q}_{1})E(u,q_{12})
   -\theta_{1}(u-2\overline{q}_{2})\theta_{2}(u-2\overline{q}_{1})p_{2}\}\Delta^{-1}\\
L_{12}(u)&=&\{-\theta_{1}(u-2\overline{q}_{1})\theta_{1}(u-2\overline{q}_{2})p_{1}
+\theta_{1}(u-2\overline{q}_{2})\theta_{1}(u-2\overline{q}_{2})E(u,q_{21})\\
& &\ \ -\theta_{1}(u-2\overline{q}_{1})\theta_{1}(u-2\overline{q}_{1})E(u,q_{12})
   +\theta_{1}(u-2\overline{q}_{2})\theta_{1}(u-2\overline{q}_{1})p_{2}\}\Delta^{-1}\\
L_{21}(u)&=&\{\theta_{2}(u-2\overline{q}_{1})\theta_{2}(u-2\overline{q}_{2})p_{1}
-\theta_{2}(u-2\overline{q}_{2})\theta_{2}(u-2\overline{q}_{2})E(u,q_{21})\\
& &\ \ +\theta_{2}(u-2\overline{q}_{1})\theta_{2}(u-2\overline{q}_{1})E(u,q_{12})
   -\theta_{2}(u-2\overline{q}_{2})\theta_{2}(u-2\overline{q}_{1})p_{2}\}\Delta^{-1}\\
L_{22}(u)&=&\{-\theta_{2}(u-2\overline{q}_{1})\theta_{1}(u-2\overline{q}_{2})p_{1}
+\theta_{2}(u-2\overline{q}_{2})\theta_{1}(u-2\overline{q}_{2})E(u,q_{21})\\
& &\ \ -\theta_{2}(u-2\overline{q}_{1})\theta_{1}(u-2\overline{q}_{1})E(u,q_{12})
   +\theta_{2}(u-2\overline{q}_{2})\theta_{1}(u-2\overline{q}_{1})p_{2}\}\Delta^{-1}\\
\Delta&=& \sigma(q_{1}-q_{2})\sigma(u-q_{1}-q_{2})
\end{eqnarray*}
\noindent and $\overline{q}_1=\frac{q_1-q_2}{2}=-\overline{q}_2$.
\noindent The Hamiltonian Eq.(2) is equal to  
\begin{eqnarray*}
H=tr(\stackrel{\sim}{L}^{2}(u))+V(u)=tr(L^{2}(u))+V(u)
\end{eqnarray*}
The motion equation can also be described in the commutator  
form 
\begin{eqnarray*}
\frac{d}{dt}L(u)=\{L(u),H\}=[L(u),M(u)]
\end{eqnarray*}
The very ``good" properties of our Lax operator defined in Eq.(7) is that 
the basical Poisson bracket of the Lax operator $L(u)$ can be written in the 
usual Poisson-Lie bracket form with a purely numerical r-matrix (i.e. 
nondynamical r-matrix). Namely,
\begin{eqnarray}
\{L_{1}(u),L_{2}(v)\}=[r_{12}(u-v),L_{1}(u)+L_{2}(v)]
\end{eqnarray}
\noindent and the numerical r-matrix $r(u)$ reads as 
\begin{eqnarray*}
r(u)=\left(\begin{array}{llll}a(u)&&&d(u)\\&b(u)&c(u)&\\
&c(u)&b(u)&\\d(u)&&&a(u)\end{array}\right)
\end{eqnarray*}
\noindent and 
\begin{eqnarray*}
& & a(u)=\frac{\theta '^{(0)}(u)}{\theta^{(0)}(u)} -\frac{\sigma '(u)}
{\sigma (u)}\ \ \ ,\ \ \ 
b(u)=\frac{\theta '^{(1)}(u)}{\theta^{(1)}(u)}-\frac{\sigma '(u)}
{\sigma (u)}\\
& & c(u)=\frac{\theta '^{(0)}(0)\theta^{(1)}(u)}
{\theta^{(0)}(u)\theta^{(1)}(0)}     \ \ \ ,\ \ \ 
d(u)=\frac{\theta '^{(0)}(0)\theta^{(0)}(u)}
{\theta^{(1)}(u)\theta^{(1)}(0)}    
\end{eqnarray*}
\noindent where $a(u),b(u),c(u),d(u)$ are all independent upon dynamical 
variable. The numerical r-matrix $r(u)$ defined in Eq.(8) satisfies the 
classical Yang-Baxter equation
\begin{eqnarray}
[r_{12}(u-v),r_{13}(u-\eta)]+[r_{12}(u-v),r_{23}(v-\eta)]
+[r_{13}(u-\eta),r_{23}(v-\eta)]=0
\end{eqnarray}
\noindent and enjoys in the antisymmetric properies 
\begin{eqnarray}
r_{12}(u)=-r_{21}(-u)
\end{eqnarray}
The standard Poisson-Lie bracket Eq.(8) of the Lax operator $L(u)$ 
and the numerical r-matrix $r(u)$ enjoying in the classical Yang-Baxter 
equation Eq.(9) and antisymmetry Eq.(10), make it possiple to construct 
the quantum theory of the CM model.
\section{The relation between two Lax representation}
In fact, the new Lax representation given by us in Eq.(7) can be 
obtained from the old one given by Krichever in Eq.(3) through a 
dynamical gauge transformation as follows
\begin{eqnarray}
L(u)=g(u)\stackrel{\sim}{L}(u)g^{-1}(u)
\end{eqnarray}
\noindent where 
\begin{eqnarray*}
g(u)=\left(\begin{array}{ll}\theta_{1}(u-2\overline{q}_{1})&-\theta_{1}(u-2\overline{q}_{2})\\
-\theta_{2}(u-2\overline{q}_{1})&\theta_{2}(u-2\overline{q}_{2})\end{array}\right)
\end{eqnarray*}
\noindent Due to the transformation Eq.(11) being dependent up the dynamical 
variable $q_{i}$ ,the classical r-matrix could not be transfered by a 
similarity transformation as that of the Lax operator in Eq.(11). Fortunately, through 
a straightforward calculation , we find that the numerical r-matrix $r(u-v)$ 
can be obtained from the dynamical one $\stackrel{\sim}{r}(u,v)$ 
as follows
\begin{eqnarray*}
r_{12}(u-v)=g_{1}(u)g_{2}(v)\stackrel{\sim}{r}_{12}(u,v)
g_{1}^{-1}(u)g_{2}^{-1}+g_{2}(v)\{g_{1}(u),\stackrel{\sim}{L}_{2}(v)\}
g_{1}^{-1}(u)g_{2}^{-1}(v)
\end{eqnarray*}
\noindent up to some matrix which does commute with $L_{1}(u)+L_{2}(v)$.
\section*{Summary}
The numerical r-matrix $r(u)$ Eq.(6) could provide a mean to construct 
a speration of variables for the CM model in the same manner as that 
in the case of the integrable magnetic chains[13]. Moreover, the r-matrix  
$r(u)$ also make it possible that one can construct the dressing 
transformation for the CM model. The dressing group of this system would be 
 an analogue to the classical limit of Sklyanin algebra[12,14]. A possible 
 way to approach quantization of the CM model would be to look for 
 a quantum version of the Poisson-Lie bracket Eq.(8). The most possible  
 candidate is the elliptic Sklyanin algebra[12,14] which satisfies 
\begin{eqnarray}
R_{12}(u-v)T_{1}(u)T_{2}(v)=T_{2}(v)T_{1}(u)R_{12}(u-v)
\end{eqnarray}
\noindent where $R_{12}(u-v)$ is the eight-vertex Baxter's R-matrix and 
satisfies the quantum Yang-Baxter equation
\begin{eqnarray}
R_{12}(u-v)R_{13}(u-\eta)R_{23}(v-\eta)=
R_{23}(v-\eta)R_{13}(u-\eta)R_{12}(u-v)
\end{eqnarray}
\noindent and the unitary condition 
\begin{eqnarray}
R_{12}(u)R_{21}(-u)=1
\end{eqnarray}
\noindent The classical numerical r-matrix $r(u)$ have the relation with 
the quantum one $R(u)$ Eq.(12) as follows 
\begin{eqnarray}
R(u)=1+wr(u)+o(w^{2})\ \ \ ,\ \ {\rm when\ \ the\ \ crossing\ \ parameter\ \ 
} w\longrightarrow 0
\end{eqnarray}

In this paper, we only consider the special case of $N=2$ for the CM model. 
However, the results can be generalized to the generic case of $2\leq N$ .
We will present the further results in the further paper.

\end{document}